\documentclass[
   final            
  ]
  {aipproc}

\layoutstyle{6x9}

\newcommand{\GeV}{~{\mathrm{GeV}}}
\newcommand{\MeV}{~{\mathrm{MeV}}}

\begin{document}

\title{Studies of heavy flavour production and the hadronic final state in high energy ep collisions}


\author{Thomas Kluge \footnote{On behalf of the H1 and ZEUS Collaborations}}{
  address={DESY, Notkestr. 85, 22607 Hamburg, Germany\\E-mail: thomas.kluge@desy.de}
}



\begin{abstract}
An extract of recent results from the H1 and ZEUS Collaborations is shown.
Various properties of quantum chromo dynamics are investigated by studying the details of the hadronic final state of high energy electron proton collisions at HERA.
The presented results include analyses of jet cross sections and single particle production such as $\gamma$ and $D$. 
Part of the measurements deal with final states involving identified heavy quarks (charm and beauty). 
\end{abstract}

\maketitle


The electron-proton storage ring HERA at DESY operates at a center of mass energy of $\sqrt s \approx 320\GeV$, allowing for precision studies of QCD by the H1 and ZEUS Collaborations.
High energy electron proton scattering can be divided in two kinematical regimes: photo production $\gamma p$ ($Q^2<1\GeV^2$) and deep-inelastic scattering DIS ($Q^2>1\GeV^2$).
In both cases pointlike interactions of the exchanged virtual photon occur, for $\gamma p$ in addition resolved processes are important, where the photon behaves as a hadron.
Heavy quarks are dominantly produced pairwise from fusion of the exchanged photon and a gluon from the proton.

\begin{figure}
  \includegraphics[width=.5\textwidth,clip]{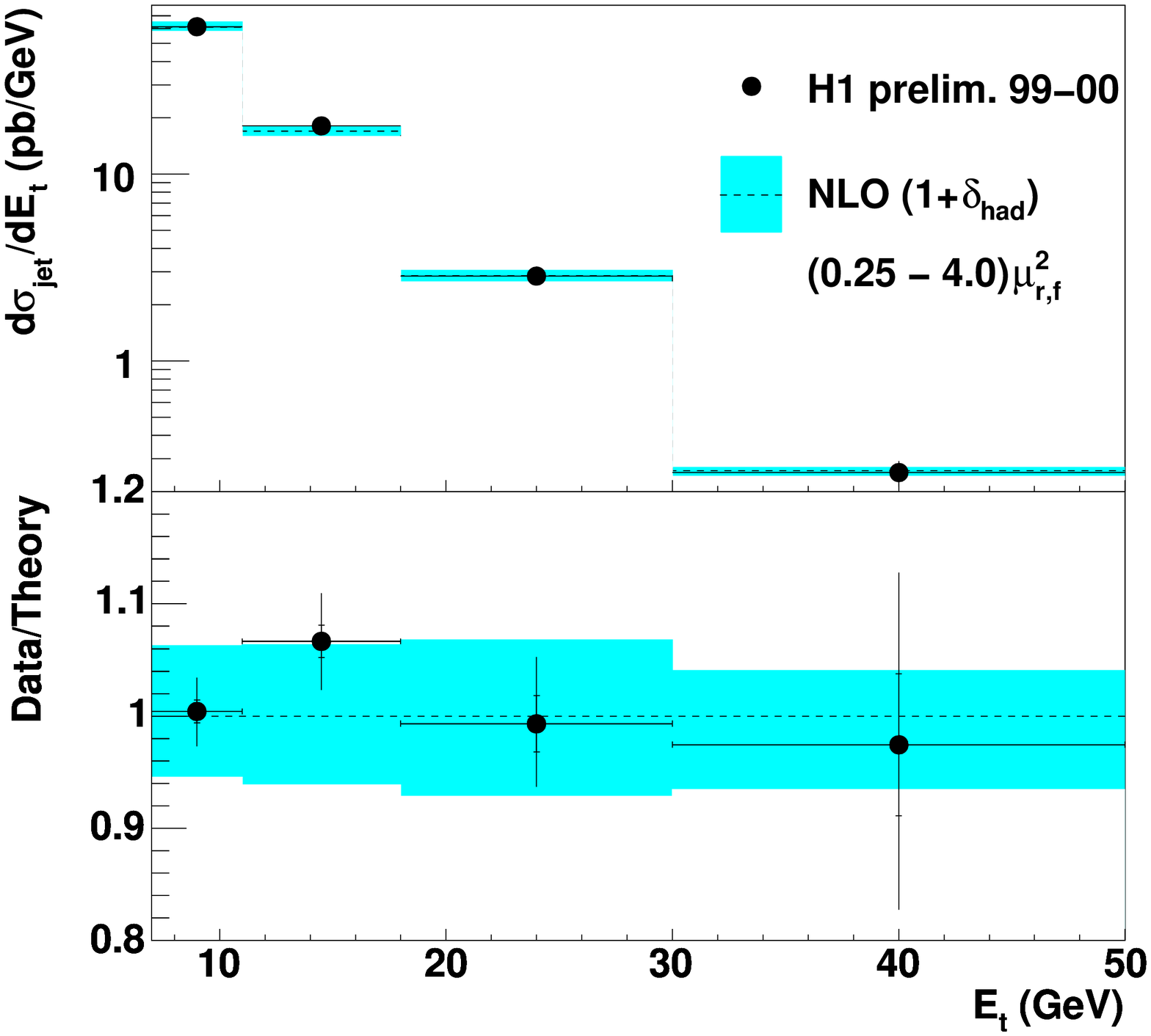}
  \includegraphics[width=.4\textwidth]{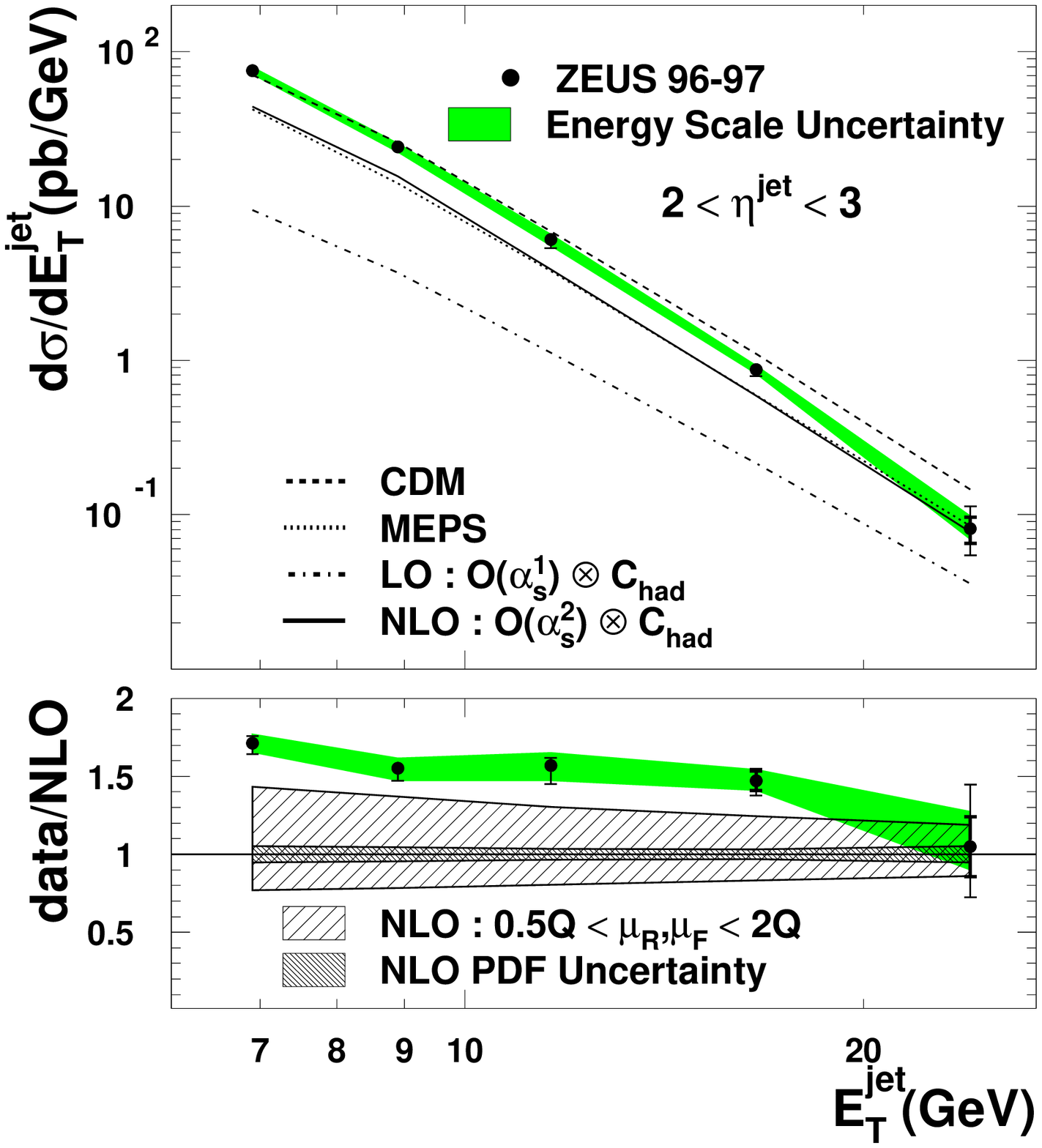}
  \label{fig1}
  \caption{Jet cross sections $d\sigma/dE_t$ compared with predictions corrected
for hadronisation effects.
The inclusive jet cross section for high $Q^2$ ($>150\GeV^2$) is shown on the left hand side plot,
 the right hand side plot shows the cross section for forward jets at lower $Q^2$ ($>25\GeV^2$).
The bands denote the theoretical uncertainty by varying the renormalization and factorization scale by a factor
of two.}
\end{figure}

Details about hard scattering and parton density functions (pdfs)  can be gained by observation of jets.
Both experimentally and theoretically well under control is the measurement \cite{incljets} of inclusive jet
cross sections at high photon virtualities ($Q^2>150\GeV^2$) and not too low transverse jet energies ($E_t>7\GeV$), shown in Fig.~\ref{fig1}.
The cross section is well described by a perturbative calculation corrected for hadronisation effects, which allows for a determination of the strong coupling constant $\alpha_s(M_Z) = 0.1197 ~\pm 0.0016\,\mathrm{(exp.)}~ ^{+0.0046}_{-0.0048}\,\mathrm{(th.)}$.

\begin{figure}
  \includegraphics[width=.4\textwidth,clip]{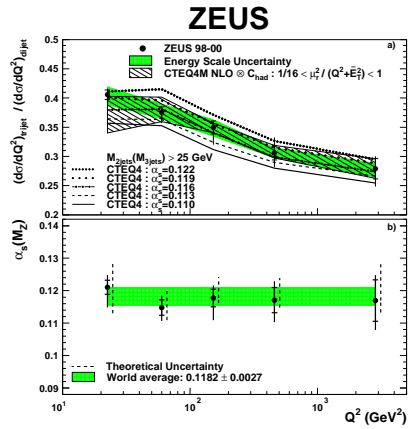}
  \label{fig2}
  \caption{The ratio of the inclusive trijet to dijet cross sections as function of $Q^2$.
The predictions of perturbative QCD in next-to-leading order using five sets of pdf are compared to the data.
The dashed error bars display theoretical uncertainties.}
\end{figure}

A more direct sensitivity to the strong coupling constant offers the observable $R_{3/2}$ \cite{Chekanov:2005ve}, the ratio of the 3- to the 2-jet cross section, which in lowest order is proportional to $\alpha_s$.
Due to the construction as a ratio of two cross sections, part of the uncertainties are expected to cancel.
The observable is well described by the theory prediction, shown in Fig.~\ref{fig2}.
An extraction of the strong coupling constant yields 
$\alpha_s(M_Z) = 0.1179 ~\pm 0.0013\,\mathrm{(stat.)}~ ^{+0.0028}_{-0.0046}\,\mathrm{(th.)}~ ^{+0.0064}_{-0.0046}\,\mathrm{(th.)}$.

The DGLAP equations are used to evolve the pdfs from a starting scale to the scale of the hard scattering.
This works well in the high $Q^2$ phase space of the before mentioned analyses, more problematic are regions of low Bj\o rken $x$ and $E_t^2 \approx Q^2$, which corresponds to jets in the forward\footnote{``Forward'' meaning in direction of the proton beam.} part of the detector.
Fig.~\ref{fig1}~(right) shows that pQCD calculations at NLO using DGLAP underestimate the data \cite{Chekanov:2005yb}.
In comparison, a prediction employing the color dipole model (CDM) does a better job, which contains terms beyond DGLAP.
However, it can  not be cleary concluded that DGLAP evolution fails, because the effects of higher perturbative orders, resp.\ the errors on them, are large.

\begin{figure}
  \includegraphics[width=.5\textwidth]{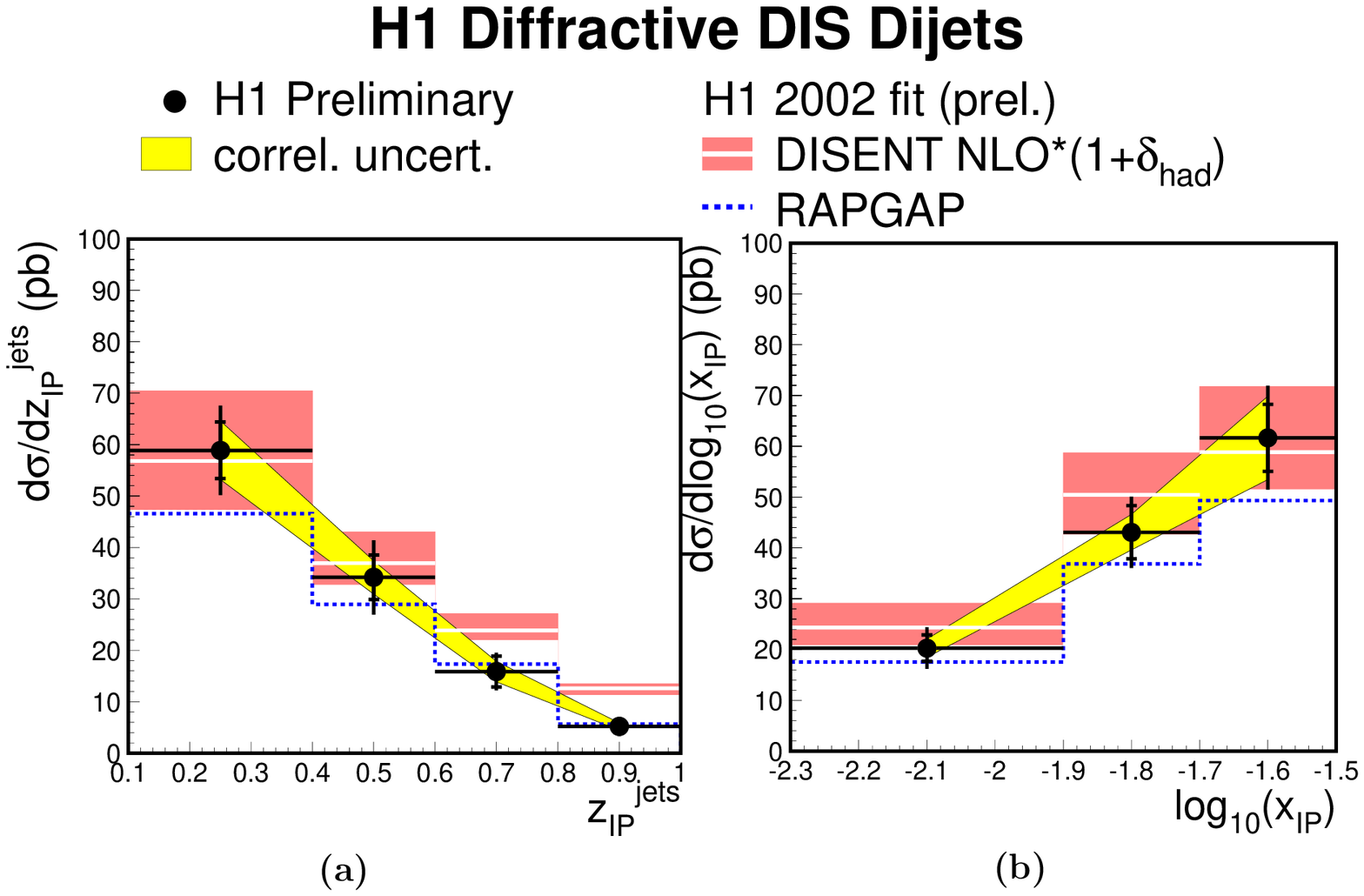}
  \includegraphics[width=.5\textwidth]{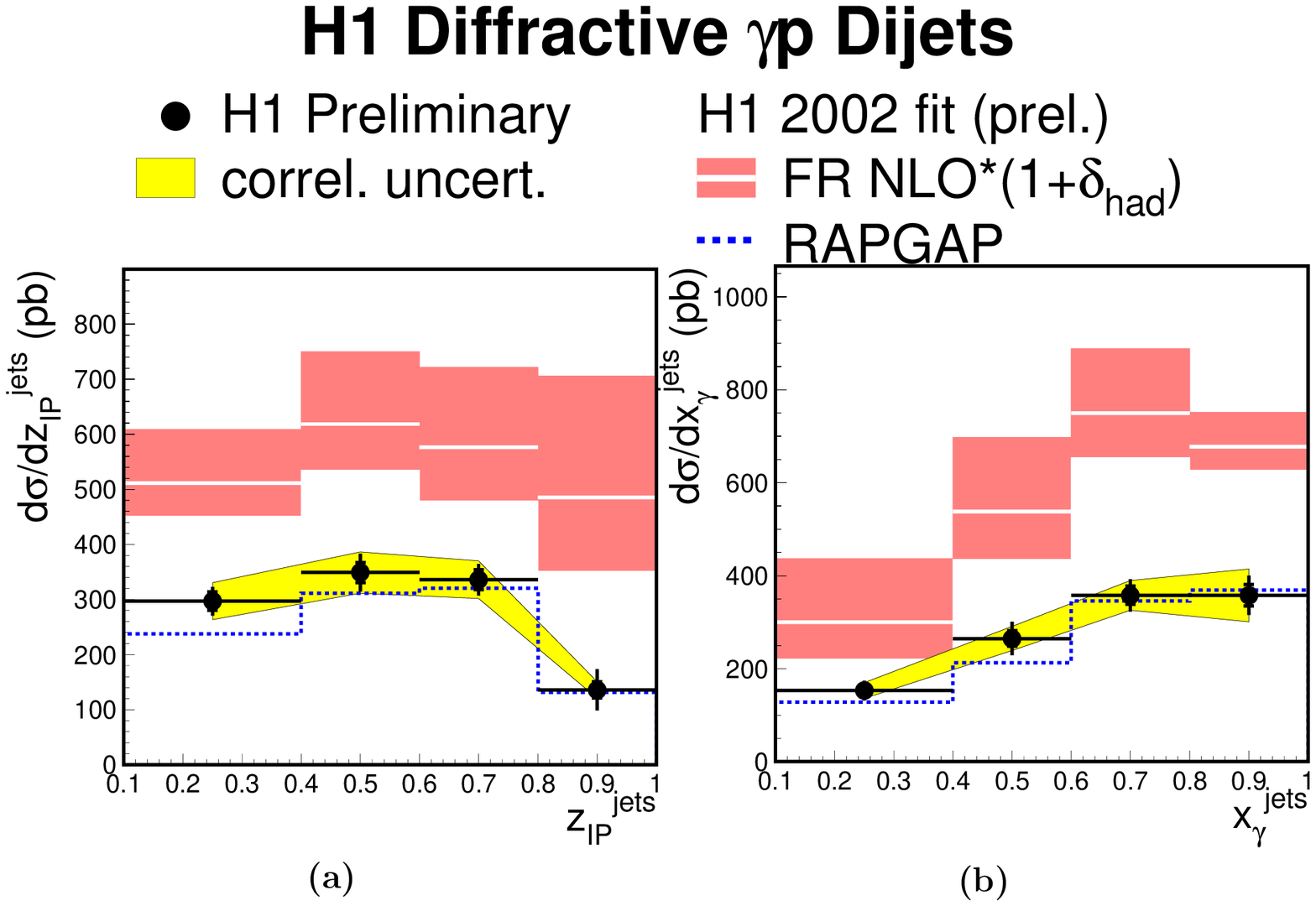}
  \label{fig3}
  \caption{Left two plots: Cross section for the diffractive production of two jets in DIS as a function of $z_P^{\mathrm jets}$
  and $\mathrm{log}_{10}(x_P)$.
  Right two plots: Cross section for the diffractive production of two jets in photo production as a function of $z_P^{\mathrm jets}$
  and $x_\gamma^{\mathrm jets}$.
}
\end{figure}

Diffractive $ep$ scattering is still a topic of great interest.
In the detector diffraction manifests as events with a rapidity gap, i.e.\ no proton dissociation occurs.
The colourless exchange with the proton can be interpreted in the pomeron picture, employing diffractive pdfs.
Such pdfs have been determined by the H1 Collaboration from inclusive data and are used to predict the  diffractive differential dijet cross section \cite{diffrdijets} in DIS, shown on Fig.~\ref{fig3}~(left).
While the overall good description suggest factorisation in DIS, the predictions for dijets in photo production  undershoot the data by far (Fig.~\ref{fig3} right).
One may conclude from this observation that factorisation is broken in $\gamma p$. 

\begin{figure}
  \includegraphics[width=.4\textwidth,angle=0,clip]{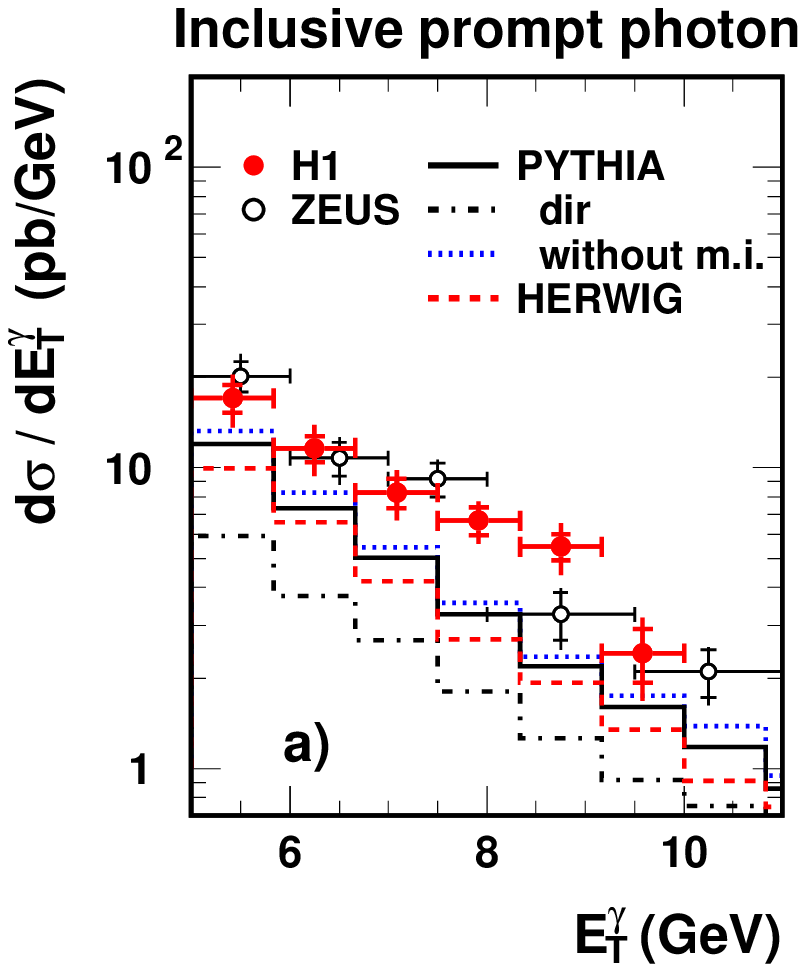}
  \includegraphics[width=.4\textwidth,angle=0,clip]{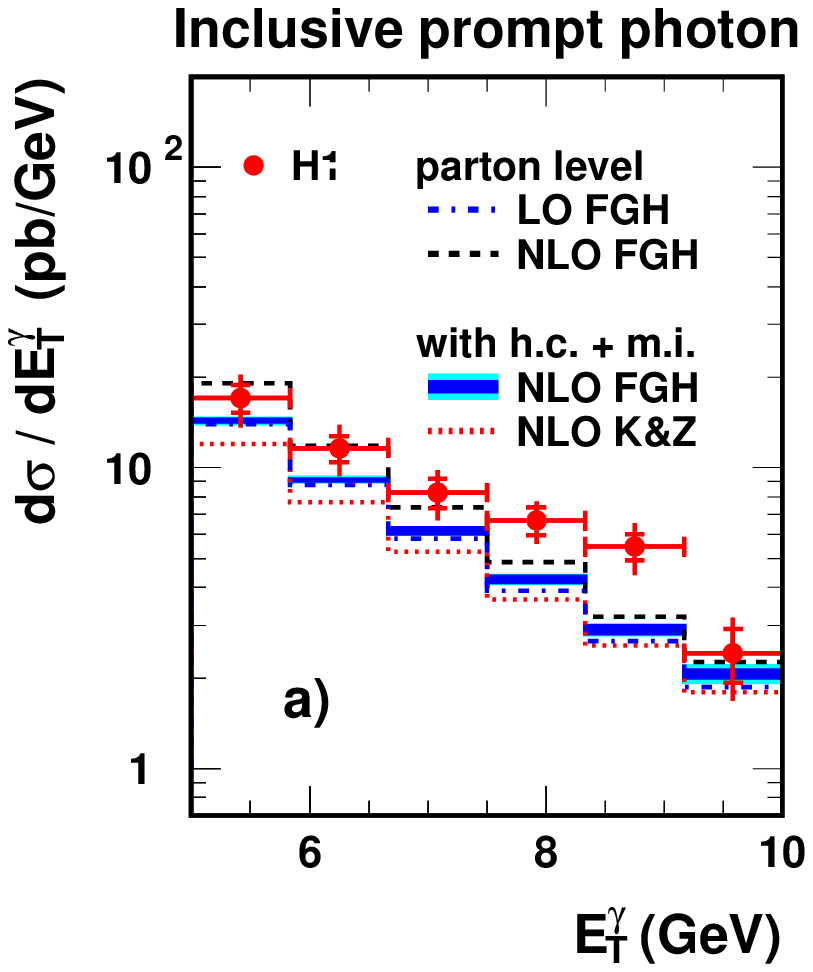}
  \label{fig5}
  \caption{Inclusive prompt photon differential cross sections in photo production kinematic range.
The data are shown as function of the transverse energy of the photon and compared to predictions 
 from HERWIG and PYTHIA including multiple interactions (left hand side).
Also shown is a comparison with NLO pQCD calculations (right hand side).
}
\end{figure}

As an alternative to the observation of jets, prompt photons emmited by quarks which participate in the hard scattering can be detected without uncertainties due to hadronisation \cite{Aktas:2004uv}.
This benefit is balanced by a reduced cross section and more difficult particle identification compared to jets. 
Fig.~\ref{fig5} demonstrates that perturbative QCD at NLO describes the shape of the transverse momentum distribution quite well, the normalisation being a bit low.
This study is complementary to QCD analyses with direct detection of jets.

\begin{figure}
  \includegraphics[width=.45\textwidth,angle=0,clip]{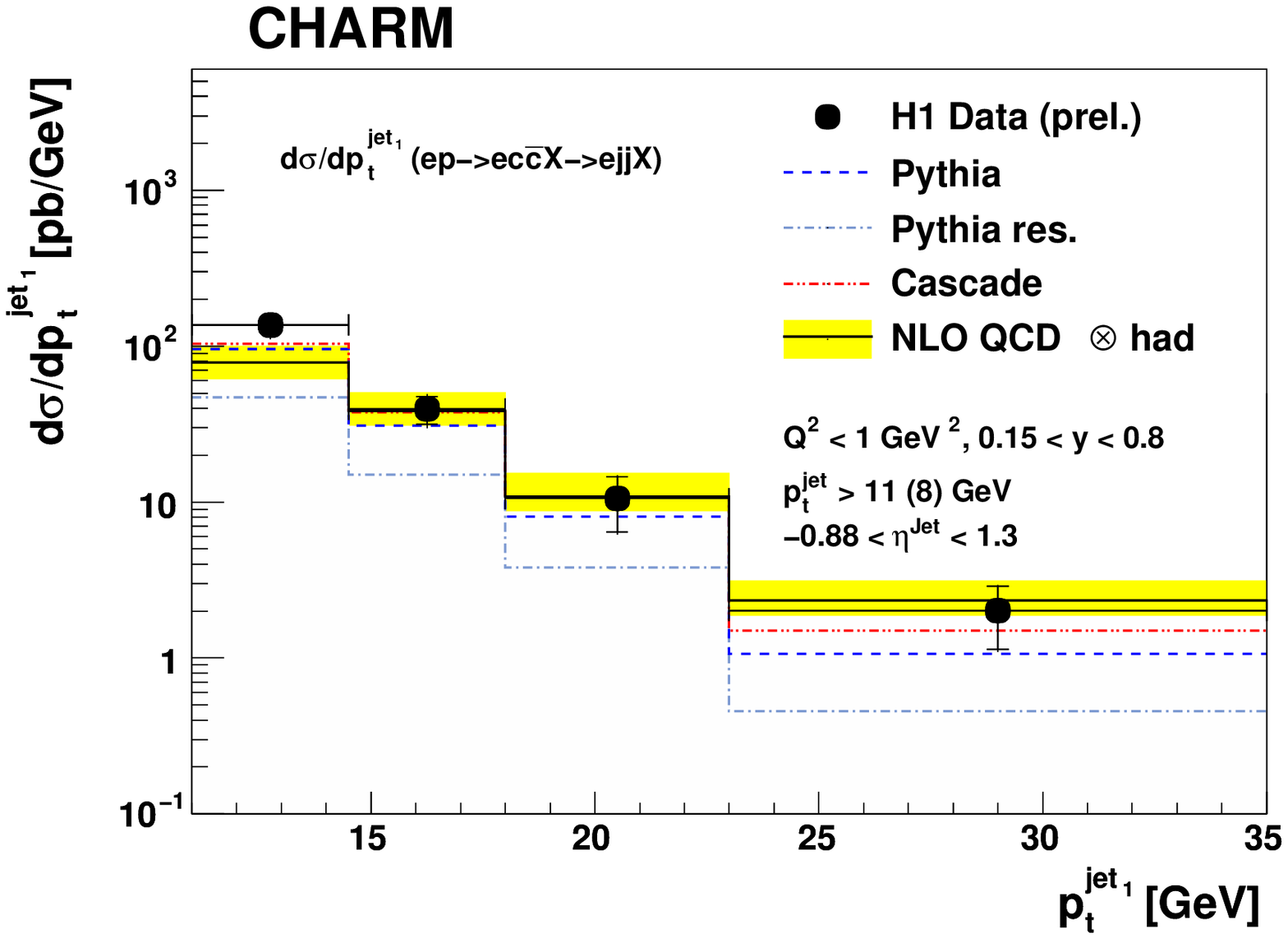}
  \includegraphics[width=.45\textwidth,angle=0,clip]{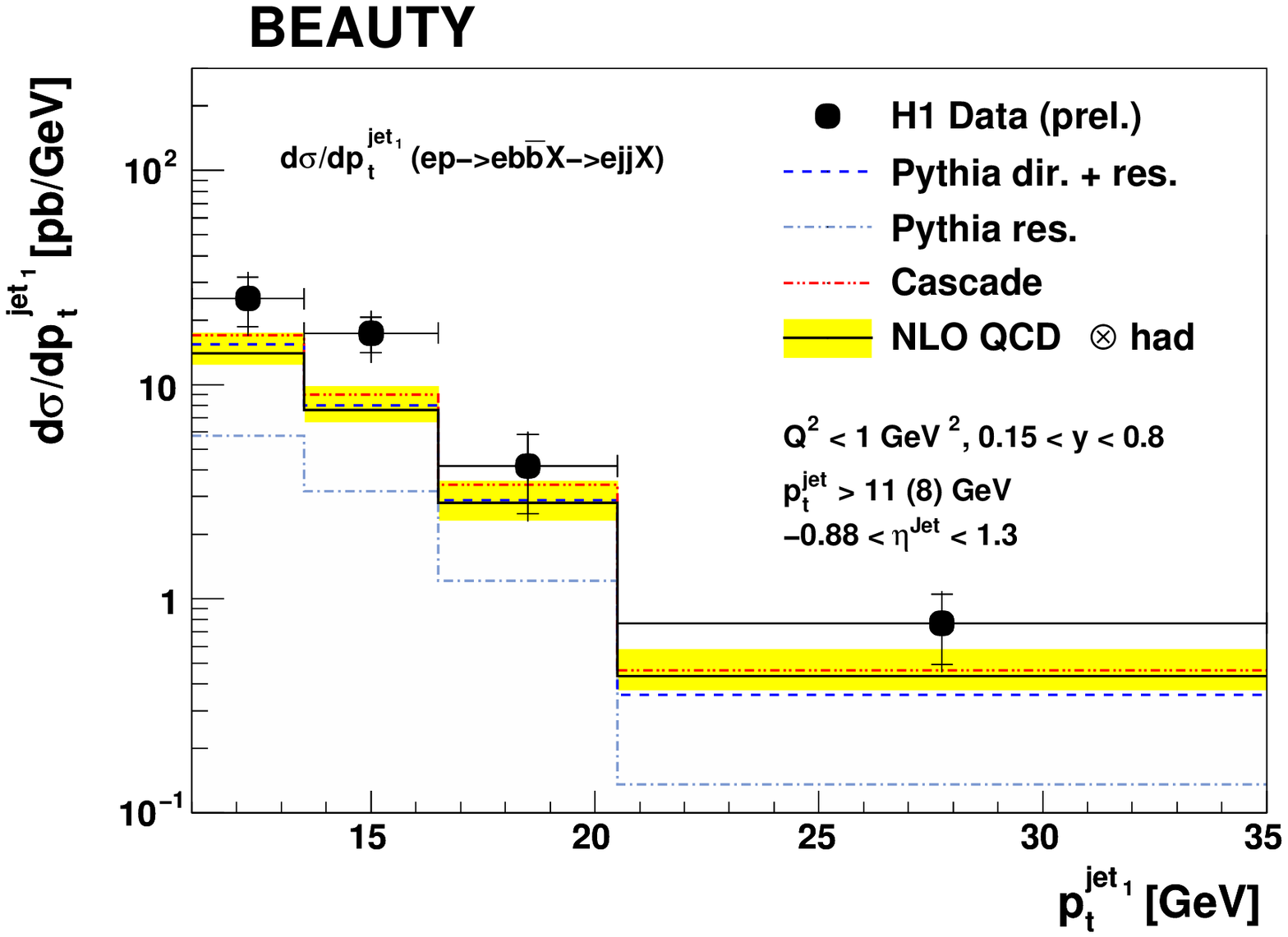}
  \label{fig4}
  \caption{Differential charm (left hand side) and beauty (right hand side) cross section as a function of the transverse
momentum of the leading jet.
The data are compared to a NLO pQCD calculation and predictions from CASCADE and PYTHIA.
The contribution in PYTHIA from resolved photon processes is shown separately.
 }
\end{figure}

Heavy quarks in final states can be identified by reconstruction of heavy mesons as $D^*$, by decay to muns or by lifetime tags. 
Fig.~\ref{fig4} shows differential cross sections for charm and beauty jets in photo production, where the fraction of heavy jets was determined by lifetime signatures.
The distributions of transverse momenta are described well by shape, the normalisation of the beauty cross section is only somewhat smaller predicted by the perturbative calculation at next-to-leading order.
This is due to improvements in experiment and theory compared to the past, where larger differences had been observed \cite{heavyjets}.

\begin{figure}
  \includegraphics[width=.45\textwidth,clip]{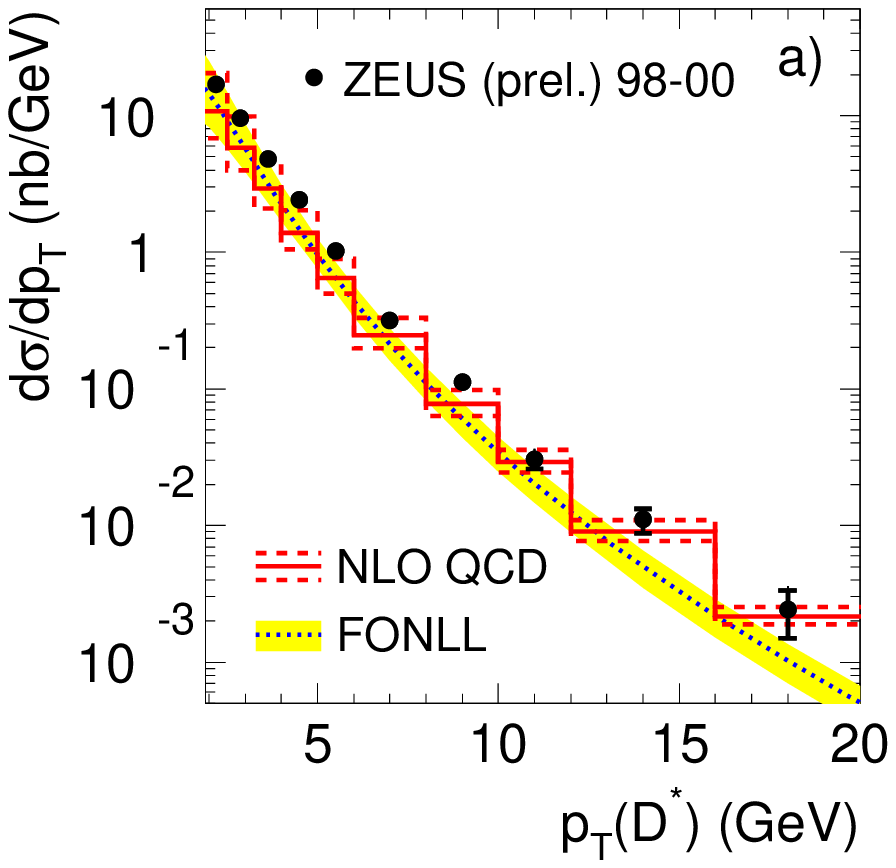}
  \includegraphics[width=.45\textwidth,clip]{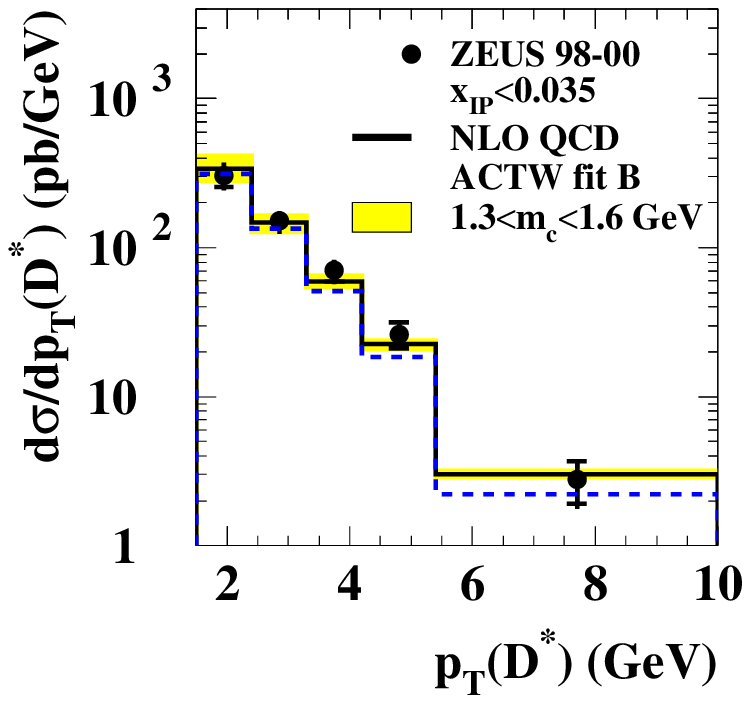}
  \label{fig6}
  \caption{Differential cross sections for inclusive $D^{*\pm}$ production as a function of the transverse energy in the kinematic range $Q^2<1\mathrm{GeV}^2$ (left). The right hand side plot shows diffractive  $D^{*\pm}$ production in DIS ($1.5<Q^2<200\mathrm{GeV}^2$).
    The data are compared to pQCD calculations at next-to-leading order (NLO).
}
\end{figure}

Detection of events containing charm can be performed by the reconstruction of a $D^*$ meson by its decay products.
Differential cross sections are shown for $\gamma p$ \cite{diffrdstar} and diffractive DIS \cite{Chekanov:2003gt} in Fig.~\ref{fig6}.
In both cases the theory predictions are able to describe the cross section.

Many observations as well as non observations of pentaquarks, bound states of five quarks, have been reported.
Searches for pentaquarks containing a strange quark decaying to $K^0_S p(\bar p)$ were performed by the H1 and ZEUS Collaborations.
In the ZEUS  analysis a signal is found near $1522\MeV$ \cite{Chekanov:2004kn}, whereby the H1 analysis \cite{penta} does not find a signal.
Exclusion limits determined by H1 do not contradict the ZEUS observation.
Hence more data, which is being recorded now during the HERA II phase, is needed to clarify this issue.
\begin{figure}
  \includegraphics[width=.37\textwidth,clip]{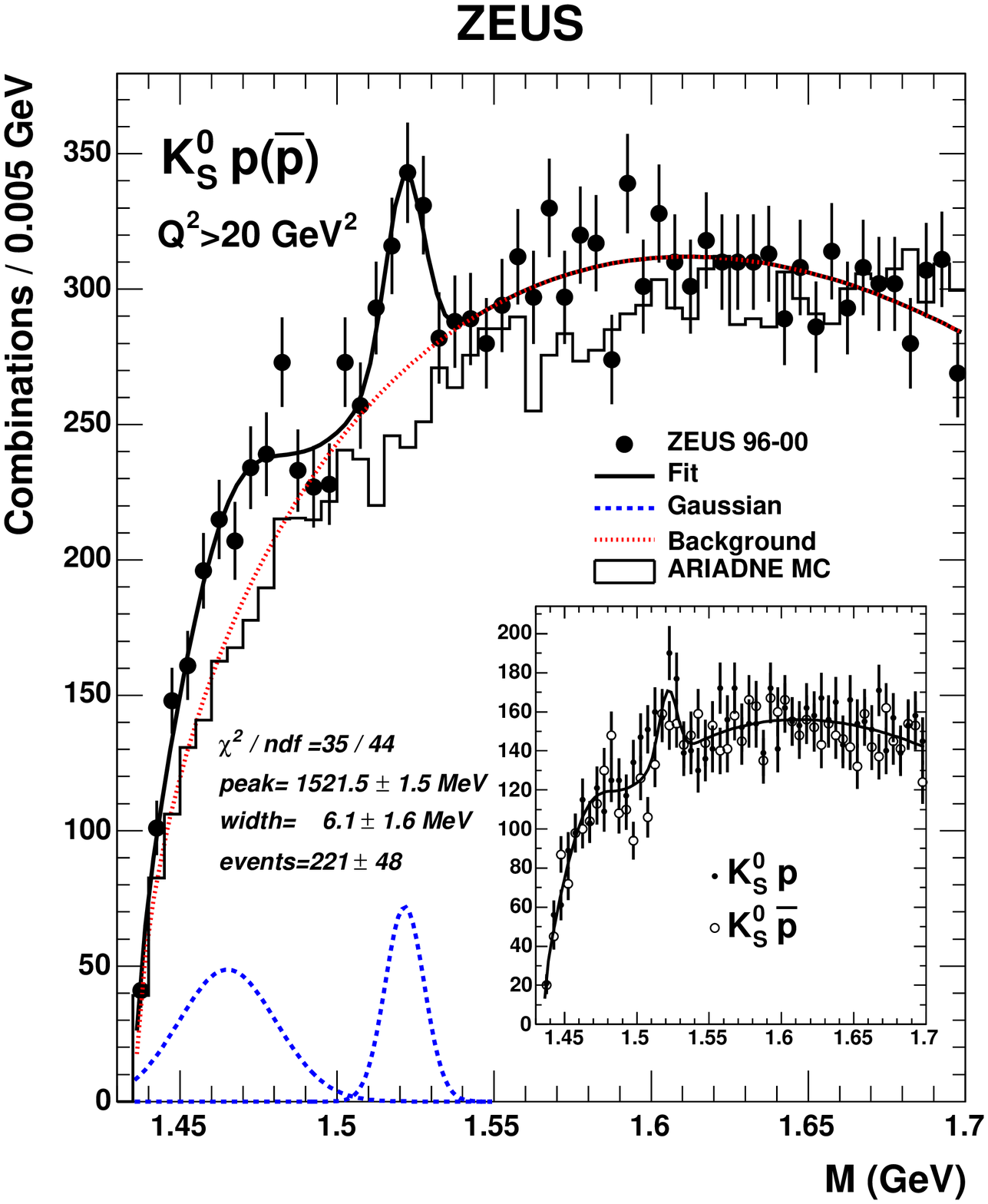}
  \includegraphics[width=.45\textwidth,clip]{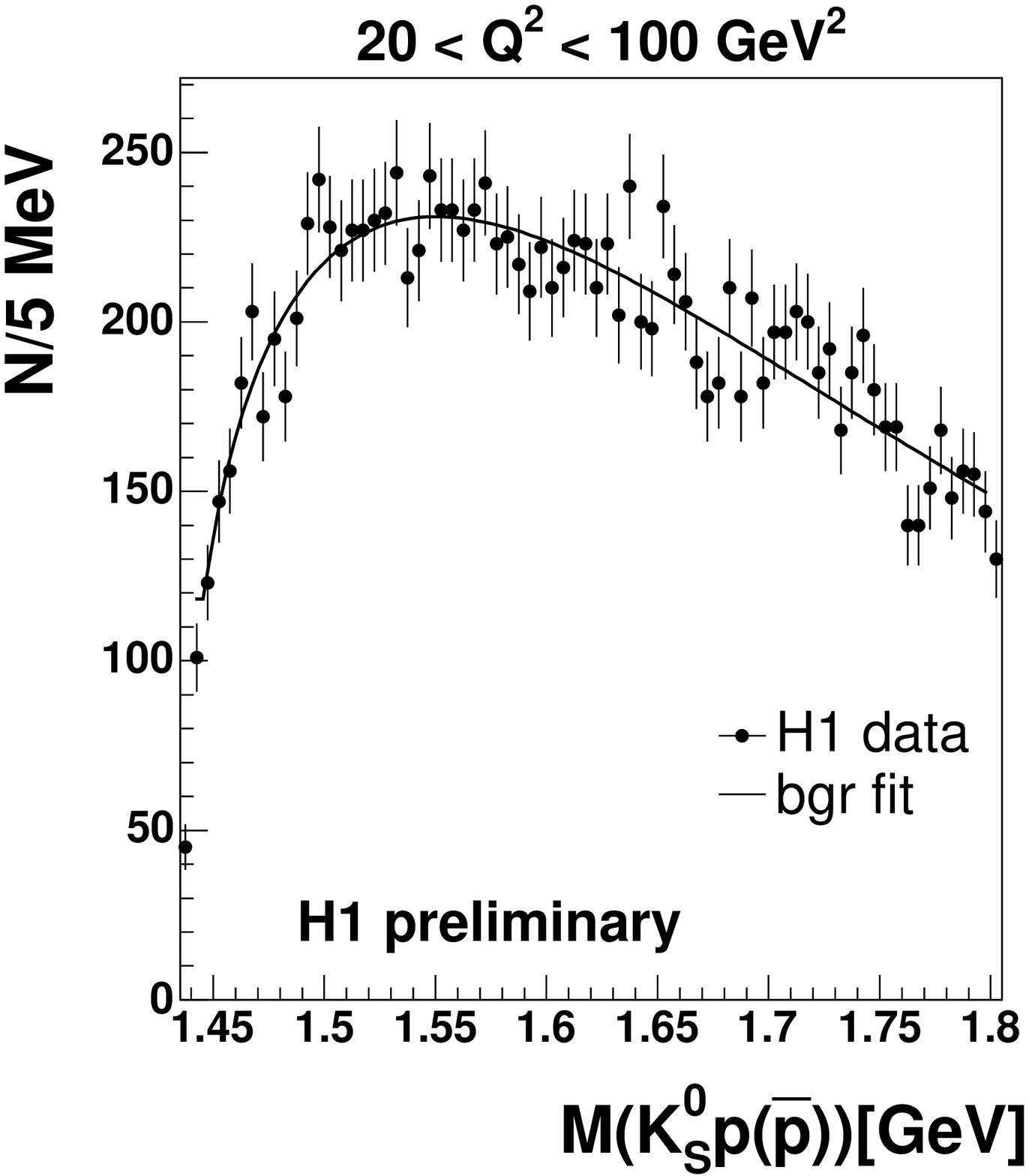}
  \label{fig7}
  \caption{Invariant mass spectrum for the $K^0_S (p\bar p)$ channel in DIS.
  Data from the ZEUS Collaboration are fitted by a background function plus two Gaussians (Left).
 The data from the H1 Collboration is fitted by a background function alone(right).}
\end{figure}





\IfFileExists{\jobname.bbl}{}
 {\typeout{}
  \typeout{******************************************}
  \typeout{** Please run "bibtex \jobname" to optain}
  \typeout{** the bibliography and then re-run LaTeX}
  \typeout{** twice to fix the references!}
  \typeout{******************************************}
  \typeout{}
 }


\end{document}